\documentclass[usenatbib,onecolumn]{mnras}
\pdfoutput=1
\usepackage{mathtext,amssymb,amsmath}
\usepackage{epsfig}
\usepackage{graphics}
\usepackage{url}
\usepackage{comment}

\usepackage{times}
\usepackage[T1]{fontenc} 
\usepackage{aecompl}

\renewcommand{\vector}[1]{\ensuremath{\mathbf{#1}}}

\newcommand{\bea}{\begin{eqnarray}}
\newcommand{\eea}{\end{eqnarray}}

\newcommand{\mdot}{\ensuremath{\dot{m}}}
\newcommand{\Msun}{\ensuremath{\,\rm M_\odot}}

\newcommand{\gfun}[1]{\ensuremath{{\rm G}\,(#1)}}

\newcommand{\pardir}[2]{\ensuremath{\frac{\partial #2}{\partial #1} }}
\newcommand{\pardirsec}[2]{\ensuremath{\frac{\partial^2 #2}{\partial #1^2} }}
\newcommand{\ppardir}[2]{\ensuremath{\frac{\partial }{\partial #1} \left( #2\right)}}

\begin{document}
\title[]{Meridional circulation and reverse advection in hot thin accretion discs}
\author[]{Pavel Abolmasov\thanks{pavel.abolmasov@gmail.com}\\
	Tuorla Observatory, Department of Physics and Astronomy, University of Turku, V\"ais\"al\"antie 20, FI-21500 Piikki\"o, Finland\\
	Sternberg Astronomical Institute, Moscow State University, Universitetsky pr. 13, Moscow 119992, Russia\\
    Kavli Institute for Theoretical Physics, University of California, Santa Barbara, CA 93106, USA }

\date{Accepted ---. Received ---; in
  original form --- }

\label{firstpage}
\pagerange{\pageref{firstpage}--\pageref{lastpage}} \pubyear{2016}
\maketitle

\begin{abstract}
In standard accretion discs, outward angular momentum transfer by viscous
forces is compensated by the inward motion of the accreting matter. 
However, the vertical structure of real accretion discs leads to meridional
circulation with comparable amplitudes of poloidal velocities. Using thin-disc
approximation, we consider different regimes of disc accretion with different
vertical viscosity scalings. We show that, while gas-pressure-dominated discs
can easily have a midplane outflow, standard thin radiation-pressure-dominated
disc is normally moving inwards at all the heights. 
However, quasi-spherical scaling for pressure
($p\propto \varpi^{-5/2}$) leads to a midplane outflow for a very broad range
of parameters. It particular, this may lead to a reversed, outward heat
advection in geometrically thick discs when the temperature decreases rapidly
enough with height. While the overall direction of heat advection depends on
the unknown details of vertical structure and viscosity mechanisms, existence
of the midplane counterflow in quasi-spherical flows is a robust result weakly
dependent on the parameters and the assumptions of the model. Future models of
thick radiatively inefficient flows should take meridional circulation into
account. 
\end{abstract}

\begin{keywords}
accretion, accretion discs -- hydrodynamics -- MHD
\end{keywords}

\section{Introduction}

One of the main assumptions of the standard thin disc approach \citep{SS73} is
locality of energy release: all the heat dissipated inside an annulus is
promptly radiated from the surface. This assumption works well as long as {\it
  (i)} the geometrical thickness of the disc is small (that ensures relatively
strong vertical gradients of the principal physical quantities) and {\it (ii)}
energy dissipation time scales are smaller than the time scales of radial
motion. Violation of the first assumption makes radiation transport more
complex, as the radiation energy density gradient is no more vertical and
diffusion coefficient variations with radius also become
significant. Violation of the second approximation due to either low
emissivity (as in radiatively inefficient advection-dominated flows, see
\citealt{NY95}) or high optical depth (as in slim discs, see
\citealt{abram88,slim}) means that the radial heat advection is important. 
Energy dissipated at some radius is then either emitted from some other
portion of the disc, or absorbed by the black hole, making the flow
radiatively inefficient. 

In steady-state accretion flows, as the mechanical energy of the flow is
transformed into heat and radiation, the matter should in general move
inwards, which makes it logical to assume that advection transports energy
inwards with the radial velocity of the flow. There are however indications
for non-trivial radial velocity profiles in certain types of accretion
discs. In particular, for isothermal, gas-pressure-dominated discs, it was
shown already by \citet{urpin84} that the radial velocity can easily have the
opposite (positive, if negative corresponds to inward motion) sign near the
equatorial plane. This finding was confirmed by numerical simulations
\citep{fromang11,stollkley,rafikov} and seems to be a usual feature for thin
disc solutions with local, Newtonian-type viscosity, expected in
protoplanetary discs around young stars. As it was shown by
\citet{rafikov}, equation (6), equatorial-plane radial velocity is sensitive
not to the vertically-integrated viscous stress, as vertically-averaged radial
velocity in the standard model, but to the local viscous stress ${\rm
  w}_{\varpi \varphi}$ and its dependence on radial (cylindrical) coordinate
$\varpi$. More specifically, if ${\rm w}_{\varpi\varphi} \varpi^2$ decreases
with $\varpi$, one should expect an equatorial-plane outflow.
In discs powered by accretion, this effect can alter the sign of the advection
term in energy equation. In contrast with protoplanetary discs, temperature in
accretion discs tends to decrease strongly with height, hence it is most
likely that the matter bearing most of the heat is moving inwards at a
lower-than-average velocity or, perhaps, even moves outwards. 

In local Newtonian-type viscosity approach, where the stress is proportional
to the strain, one should also take into account the angular
velocity dependence on vertical coordinate. 
Even though for thin discs angular frequency changes only slightly with
height,
its second vertical derivative is of the same order as the second order radial
derivative, and thus may contribute significantly to angular
momentum transport. This effect makes the angular momentum transport problem
more complicated, but is straightforward to account for in certain assumptions
about the vertical structure. 

In gas-dominated parts of a standard disc, pressure decreases with radius
rapidly enough to suggest a midplane outflow, if viscous stress scales
linearly with pressure (the approach we will hereafter call local
$\alpha$-assumption). However, the effects of advection are very small in
gas-dominated discs. 
On the other hand, simple dimensional arguments suggest that, whenever
advection is strong and accretion disc becomes thick, the speed of sound
scales with virial velocity $c_{\rm s} \propto \varpi^{-1/2}$, and pressure
becomes a very steep function of radius, $p\propto \varpi^{-5/2}$, which is
strongly suggestive for existence of a midplane outflow in any type of a thick
disc. 

It is of course difficult to extrapolate the results obtained in thin-disc
approximation to geometrically thick flows. However, it is possible to make
estimates for the midplane radial velocity which do not rely much on the
assumption of disc thickness. 

The goals of this paper are to estimate the vertical rotational structure and
two-dimensional angular momentum transport in thin accretion discs where the
vertical and radial structure may be decoupled, and to use this approach to
estimate the direction of heat advection in different kinds of accretion discs
and flows. In Section~\ref{sec:gene}, we formulate the problem and introduce
the notation used throughout the paper. Then, in Section~\ref{sec:rot}, we
calculate the vertical structure of the angular frequency field. In
Section~\ref{sec:angmo} we use different local viscosity models to recover the
expected angular momentum transfer and the poloidal velocity field. In
Section~\ref{sec:heat}, we show that existence of a midplane outflow in an
accretion disc often leads to outward advection of the energy stored as
trapped radiation. In Section~\ref{sec:disc}, we discuss the validity of the
assumptions used and consider the { possible extensions of the model. }

\section{Basic assumptions and notation}\label{sec:gene}

We will use cylindrical coordinates $z$ and $\varpi$, where $z$ is the height
above the equatorial plane, and $\varpi$ is the distance towards the axis. We
assume stationarity and axisymmetry. Poloidal velocity components are always
negligible in comparison with the azimuthal component $v_\varphi = \varpi
\Omega$. We assume that above the finite height of $z=H$, density and pressure
in the accretion disc become zero. We will also neglect all the terms of the
order $\left(H/\varpi\right)^2$ and higher in dynamical equations, and assume
that, for every physical quantity $f=f(\varpi, z)$, it is possible to separate
the variables as $f=f_\varpi(\varpi) f_z(z/H(\varpi))$. Besides, all the
radial dependences are assumed smooth and approximated by power laws using
notation $f_\varpi \propto \varpi^{-\Gamma_f}$. Rotation velocity $\Omega$ is
assumed close to Keplerian $\Omega_{\rm K} = \sqrt{GM} \varpi^{-3/2}$ with
accuracy $\sim \left(H/\varpi\right)^2$. As the angular momentum transfer
equation for Newtonian-type viscosity contains second vertical derivatives of
angular frequency $\partial^2\Omega/\partial z^2 \sim \Omega / H^2$, vertical
variations of rotation velocity, though negligibly small by themselves,
produce { a significant contribution to the angular momentum transport equation.}

The set of assumptions we use is supposed to reproduce the main properties of the standard $\alpha$-disc after vertical integration.
Strong deviations from the standard model may arise in the models where the viscous stress does not zero at the surface of the disc that implies angular momentum exchange with some ambient medium, such as wind or corona. We will assume that the viscous stress zeroes at the surface. 

Dynamical equations we use below are derived from the stationary momentum (Navier-Stokes) equation
\begin{equation}\label{E:Euler}
\rho (\vector{v}\nabla) \vector{v} = - \nabla p - \rho \nabla \Phi - \nabla \mathrm{w},
\end{equation}
where ${\rm w}$ is the second-rank viscous stress tensor, components of which are assumed small with respect to pressure. We also suggest that, as poloidal velocity is everywhere smaller than azimuthal, only ${\rm w}_{r\varphi}={\rm w}_{\varphi r}$ and ${\rm w}_{z\varphi}={\rm w}_{\varphi z}$ should be taken into account. Together, these two assumptions imply that only the $\varphi$ component of equation~(\ref{E:Euler}) is affected by viscosity. 

Standard disc model uses all these assumptions, but only for vertically-integrated quantities, as the viscosity processes operating in real accretion discs are presumably determined by the characteristic spatial scales of accretion disc thickness. When viscosity is determined by microphysics (ionic or radiative viscosity), kinetic theory \citep{kinetics} predicts Newtonian-viscosity stress-strain relation (expressed in general Cartesian coordinates $i$, $j$, and $k$) of the form
\begin{equation}\label{E:stressstrain}
\displaystyle {\rm w}_{ik} = \left( \pardir{x_i}{v_k}+ \pardir{x_k}{v_i} - \frac{2}{3} \delta_{ik} \pardir{x_j}{v_j}\right) \nu \rho \simeq \left( \pardir{x_i}{v_k}+ \pardir{x_k}{v_i} \right) \nu \rho,
\end{equation}
where $\nu$ is viscosity coefficient determined by the characteristic spatial $l_{\rm t}$ and velocity $v_{\rm t}$ scales of the transfer process. In the case of turbulent viscosity, $l_{\rm t}$ is characteristic mixing length and $v_{\rm t}$ is characteristic turbulent viscosity. If magnetic fields dominate momentum transfer, these quantities may be understood as characteristic magnetic loop size and Alfv\'{e}n velocity. For meso-scale processes with $l_{\rm t} \sim H$, the stress-strain scaling should be more complex than in Newtonian approximation and may involve finite differences or spatially-averaged partial derivatives instead, but the direction of angular momentum transport should remain approximately the same, while the amplitude of the stress may vary. Thus equation~(\ref{E:stressstrain}) should still work well unless the transfer processes become anisotropic (such as anisotropic turbulent velocity field considered by~\citealt{wasiutynski}). In the latter case, $\nu$ is no more scalar, and the direction of viscous angular momentum transfer can differ from the direction of velocity gradient. This results in additional terms in viscous stress  (see, for instance,\citealt{tayler73} and \citealt{tassoul} chapter 8), that may change completely the angular momentum transfer in accretion discs if anisotropy is strong, as in convection-dominated flows \citep{QG00, igu02}. We will not consider the effects of anisotropic momentum transfer in this paper, except a brief discussion in Section~\ref{sec:disc:aniso}. 

For our case, the velocity field may be approximated as a differential rotation field, and
\begin{equation}
\frac{{\rm w}_{ik}}{\varpi \nu \rho} = \left(\vector{e}^\varphi_i \vector{e}^r_k + \vector{e}^r_i \vector{e}^\varphi_k\right) \pardir{\varpi}{\Omega} + \left(\vector{e}^\varphi_i \vector{e}^z_k + \vector{e}^z_i \vector{e}^\varphi_k\right) \pardir{z}{\Omega},
\end{equation}
where expression $\vector{e}^b_i$, for $b=\varpi$, $z$, or $\varphi$, is the $i$-th component of the corresponding (radial, vertical, or azimuthal) unit coordinate vector of the cylindrical coordinate frame. 
The relevant, azimuthal, component of the stress tensor gradient then becomes
\begin{equation}\label{E:wik}
\displaystyle \vector{e}^\varphi_i \nabla_k {\rm {\rm w}_{ik}} = \frac{1}{\varpi^2} \ppardir{\varpi}{\nu \rho \varpi^3 \pardir{\varpi}{\Omega}} + \varpi \ppardir{z}{\nu \rho \pardir{z}{\Omega}}.
\end{equation}
This expression is quite general, and the local $\alpha$-viscosity scaling may be reproduced by substituting $\nu \rho =\alpha p / |\nabla \Omega| \propto \varpi p/\Omega$. 

The angular momentum transfer equation is obtained as the azimuthal component of~(\ref{E:Euler})
\begin{equation}\label{E:angmo}
\frac{1}{\varpi}\rho v_\varpi \ppardir{\varpi}{\Omega \varpi^2} = e^\varphi_i \nabla_k {\rm {\rm w}_{ik}},
\end{equation}
where the right-hand side will be assumed to be given by~(\ref{E:wik}). In the
left-hand side, we neglected the term proportional to $v_z$, as
it is by a factor $\displaystyle \sim \left(\frac{H}{\varpi}\right)^2$ smaller
than the radial term.

The other two independent components of the momentum equation will be used to
solve for the vertical profiles of pressure and angular frequency. As poloidal
velocity components are small, and vertical gradients prevail over radial,
vertical component of momentum equation is simply hydrostatic equilibrium
condition
\begin{equation}\label{E:hyst}
\pardir{z}{p}=-\Omega_{\rm K}^2 z \rho.
\end{equation}
As the last of the three independent dynamical equations, it is convenient to use azimuthal component of the curl of equation~(\ref{E:Euler}). Neglecting poloidal velocity components, it can be written as (see also~\citealt{tassoul}, equation 27 in chapter 4)
\begin{equation}\label{E:oz}
\varpi \ppardir{z}{\Omega^2} = \ppardir{z}{\frac{1}{\rho}}\pardir{\varpi}{p} -  \ppardir{\varpi}{\frac{1}{\rho}}\pardir{z}{p}.
\end{equation}
Using equations~(\ref{E:hyst}) and (\ref{E:oz}) allows to calculate the vertical profiles of pressure and angular frequency given the vertical density profile. To recover $\rho$ as a function of $z$, an additional constraint is needed, that could be effective equation of state or energy equation. Both approaches will be considered in Section~\ref{sec:rot}. Note that existence of a global equation of state $p=p(\rho)$ makes angular frequency constant with height. 

\section{Rotation profile}\label{sec:rot}

\subsection{Analytical approximation}\label{sec:aansatz}

It is convenient to parameterize the vertical density profile as
\begin{equation}\label{E:rhoansatz}
\rho = \rho_{\rm c}(\varpi) \left( 1-\xi^2\right)^a,
\end{equation}
where $\xi = z/H$, and $a$ is a free parameter. Hydrostatic
equation~(\ref{E:hyst}) is then easy to integrate obtaining the
vertical pressure profile
\begin{equation}\label{E:verpressure}
p = p_{\rm c}(\varpi) \left( 1-\xi^2\right)^{a+1},
\end{equation}
where
\begin{equation}\label{E:pc}
p_{\rm c} = \frac{\Omega_{\rm K}^2 H^2}{2(a+1)}\rho_{\rm c}.
\end{equation}

In the barotropic case, our ansatz corresponds to a polytrope, and $a$ is the
polytropic index, $p=\rho^{1+1/a}$. In other cases it can be viewed as an
effective polytropic index. In particular, { isentropic}
radiation-pressure-supported disc with $p\propto \rho^{4/3}$ is expected to
have $a=3$. As we will see below, $a=3$ is a good approximation even for
gas-dominated solutions. On the other hand, more sophisticated models of
vertical structure \citep{KS98,SSZ} predict $a\simeq 3$ for gas-dominated
discs and $a\simeq 1$ for radiation-pressure-dominated discs with
convection. Hence, different values of $a$ should be considered. 

In such a parameterization, equation~(\ref{E:oz}) takes the form
\begin{equation}\label{E:ozga}
\pardir{z}{\Omega} = \frac{a\Gamma_p-(a+1)\Gamma_\rho}{2(a+1)} \frac{z}{\varpi^2}  \Omega. 
\end{equation}
Quantities $\Gamma_{p}$ and $\Gamma_\rho$ are logarithmic partial derivatives
of central pressure and density, respectively, with respect to $\varpi$. 
For the second derivative, neglecting the higher-order terms in $z/\varpi$, 
\begin{equation}\label{E:ozzga}
\pardirsec{z}{\Omega} \simeq \frac{a\Gamma_p-(a+1)\Gamma_\rho}{2(a+1)} \frac{1}{\varpi^2} \Omega, 
\end{equation}
constant with height as long as the effective polytrope index $a$ is constant. Integration of (\ref{E:ozga}) yields the following, Gaussian angular velocity profile
\begin{equation}\label{E:oga}
\displaystyle \Omega = \Omega_{\rm c} e^{\frac{a\Gamma_p-(a+1)\Gamma_\rho}{4(a+1)} \frac{z^2}{\varpi^2}},  
\end{equation}
where $\Omega_{\rm c}$ is the midplane rotational frequency, equal to Keplerian with an accuracy $\sim (H/\varpi)^2$. 
Evidently, the sign of $a\Gamma_p-(a+1)\Gamma_\rho$ defines the shape of the isostrophes (surfaces of constant $\Omega$) in the disc near the equatorial plane. Generally, the sign of this combination is negative, hence the isostrophes are convex. 

Important special case not covered by the parameterization used here but easily derivable as the limit $a\to \infty$ is isothermal disc. Density profile has Gaussian shape in this case, hence there is no surface where density approaches zero. Angular frequency derivative becomes
\begin{equation}\label{E:ozga:iso}
\left.\pardir{z}{\Omega}\right|_{\rm isothermal} = \left(\Gamma_p-\Gamma_\rho\right) \frac{z}{\varpi^2} \Omega. 
\end{equation}
If pressure decreases with the radial coordinate more rapidly than density, isostrophes are convex, and vertical shear transports angular momentum upwards. 

\subsection{Numerical solution}\label{sec:num}

It is possible to use a more physical approach by replacing the effective
polytrope with assumptions about heat release, radiative transfer and
opacity. Vertical structure of thin accretion discs was calculated in a number
of papers like \citet{KS98,SSZ}, but here we also track the vertical profile
of angular velocity $\Omega(z)$. To check the validity of the
effective-polytrope approach for predicting rotational structure and angular
momentum transport, we considered the particular case of viscous stress
independent of height. This is a reasonable assumption if effective viscosity
is provided by magnetic fields (see below Section \ref{sec:angmo}). Details of
the vertical structure model are given in
Appendix~\ref{sec:app:vert}. Dimensionless mass accretion rate is $\dot m
=1$. We normalize the mass accretion rate as $\displaystyle \dot m = \frac{\dot M c^2}{L_{\rm
    Edd}} = \frac{\dot M \varkappa_{\rm T} c}{4\pi GM}$, where
$\varkappa_{\rm T}\simeq 0.34{\rm\, cm^2 g^{-1}}$ is Thomson electron
scattering opacity. Same definitions are used in
Appendix~\ref{sec:app:vert}.

\begin{figure*}
\includegraphics[width=\columnwidth]{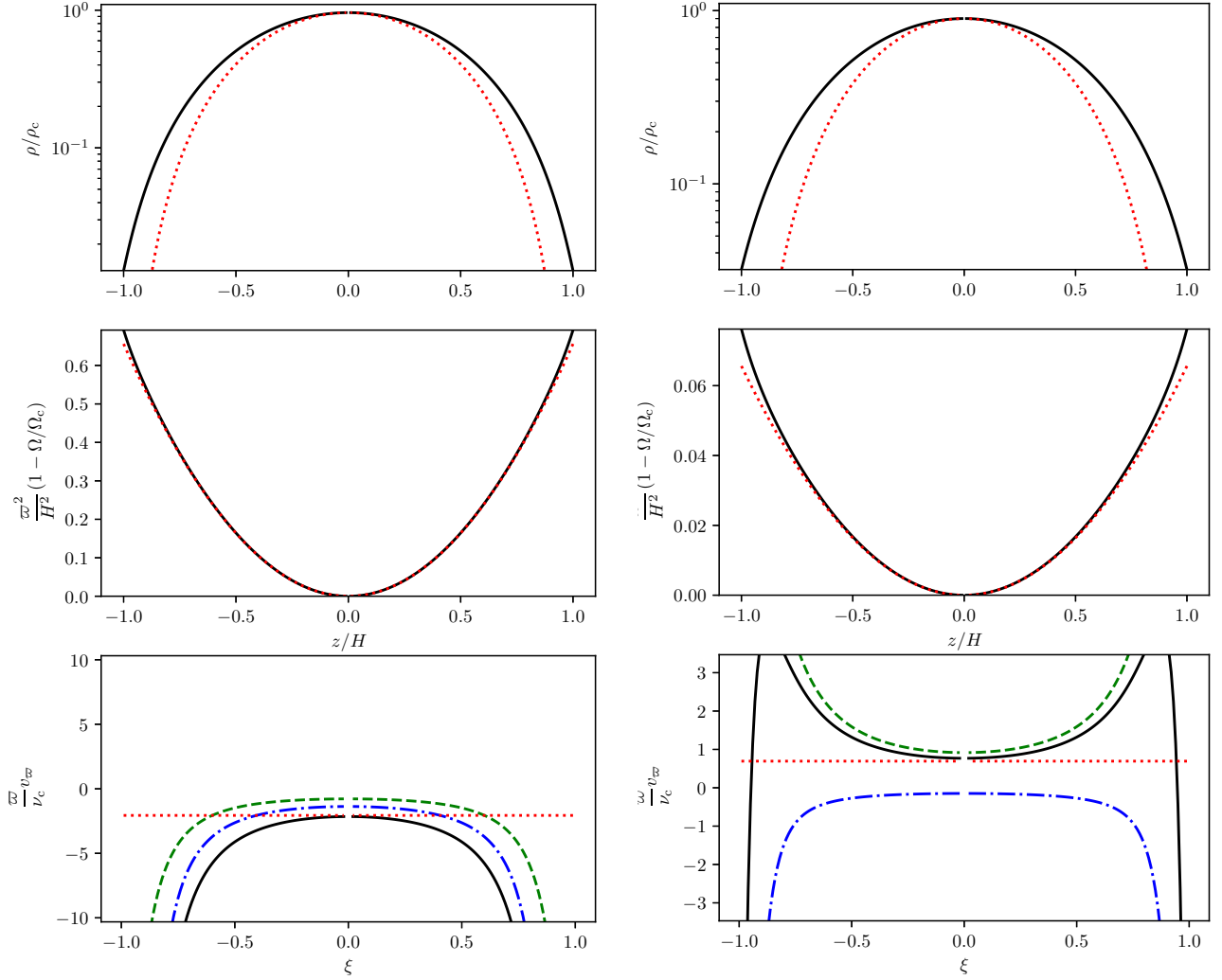}
 \caption{Vertical structure of constant-stress accretion in
   radiation-pressure- (left panels) and gas-pressure-dominated
   regimes. Dimensionless mass accretion rate is $\dot m =1$ for both cases,
   distance from the central object $\varpi = 100$ and
   $1000\,\frac{GM}{c^2}$. Upper, middle and lower panels show density
   profiles, angular frequency variations, and radial velocities,
   respectively. Velocities are normalized over characteristic values. 
   Red dotted lines are predictions of the $a=3$
   effective-polytrope model. Green dashed and blue dot-dashed lines in the
   lower panels correspond to radial and vertical shear contributions to
   radial velocity (shown in black solid). }\label{fig:verstre:ab}       
\end{figure*}

In Fig.~\ref{fig:verstre:ab}, we show the sample { vertical structure of
an accretion disc at different radii,} corresponding to different
cases of standard disc accretion, particularly to electron-scattering,
radiation- and gas-pressure-dominated zones A and B. To calculate $\Omega$, we
integrated equation~(\ref{E:oz}) together with the vertical structure
equations, assuming power-law dependences on radius for all the midplane
quantities. If the vertical structure is described by universal functions of
$z/H$, the variation of disc thickness $H$ with radius does not affect
equation~(\ref{E:oz}) as the terms containing ${\rm d}H/{\rm d}\varpi$ cancel
out:
\begin{equation}
\displaystyle \pardir{\xi}{\ln \Omega} = \frac{1}{2\varpi^2 \Omega^2} \frac{p}{\rho} \left( - \Gamma_p \pardir{\xi}{\ln \rho}+\pardir{\xi}{\ln \rho }\pardir{\xi}{\ln p} \Gamma_H + \Gamma_\rho \pardir{\xi}{\ln p} - \pardir{\xi}{\ln \rho }\pardir{\xi}{\ln p} \Gamma_H\right) = \frac{1}{2\varpi^2 \Omega^2} \frac{p}{\rho} \left( \Gamma_\rho \pardir{\xi}{\ln p}- \Gamma_p \pardir{\xi}{\ln \rho} \right).
\end{equation}

As it can be seen in Fig.~\ref{fig:verstre:ab}, effective polytrope with $a=3$ is a reasonable estimate for the disc solution with $\nu \rho = const$ and, probably, for real discs whenever vertical convection does not play important role. 

\section{Angular momentum transport}\label{sec:angmo}

Angular momentum transport is governed by two terms, advective and viscous
(the left- and right-hand sides of equation~\ref{E:angmo}), that should
exactly cancel in stationary case, as angular momentum is a conserved
quantity. This allows to estimate the radial velocity as a function of height
using equation~(\ref{E:angmo}). For a thin disc, it is safe to replace
$\Omega$ with $\Omega_{\rm K}$ (that is of course not true for second
derivatives with height). 

\subsection{Barotropic rotation}

This is the case when for some reason the matter of the disc has a global barotropic equation of state. 
As we will see later, some of the basic properties of barotropic models may be extrapolated to the more general case when $\Omega$ variation with height is allowed for. 

As it is stated by Poincar\'{e}-Wavre's theorem (see \citealt{tassoul},
section 4.3), angular frequency in a stationary barotropic differentially
rotating flow is a function of the cylindrical radial coordinate only. If
$\Omega=\Omega(\varpi)$, the second term in the right-hand side of
equation~(\ref{E:wik}) becomes zero. For Keplerian rotation, this implies
\begin{equation}\label{E:barwik}
\displaystyle \left(e^\varphi_i \nabla_k {\rm {\rm w}_{ik}}\right)_{\mbox{barotropic}} = -\frac{1}{\varpi^2}\ppardir{\varpi}{{\rm w}_{r\varphi} \varpi^2}.
\end{equation}
As it is easy to see, the sign of the whole expression, and, subsequently, the
radial velocity in the midplane is defined by the radial dependence of the
local viscous stress: if ${\rm {\rm w}_{\varpi \varphi}}\varpi^2$ decreases
with radius, there will be a counterflow in the midplane. If rotation is close
to barotropic, this quantity is also representative for mean radial velocity.  
In particular, local-alpha approach predicts
\begin{equation}\label{E:barvr}
v_\varpi = - \frac{2\alpha}{ \Omega \varpi^2 \rho}\ppardir{\varpi}{p \varpi^2}.
\end{equation}
One of the consequences is that if the variables are separable and equation of
state is barotropic (or pseudo-barotropic), the sign of radial velocity does
not change with vertical coordinate. The sign of the radial velocity is
defined by pressure dependence on $\varpi$. If $\Gamma_p < -2$, the sign
of $v_\varpi$ is positive, and accretion is
impossible. Expression~(\ref{E:barvr}) does not rely on the thinness of the
disc as long as $z\ll \varpi$, hence it is also valid, for instance, for the
midplane of a thick barotropic torus. For a geometrically thick, locally
viscous barotropic solution it states that a midplane outflow will be present
if the pressure decreases rapidly enough with distance. From dimensional
arguments, the radial pressure dependence in thick discs with $H\propto
\varpi$ may be shown to depend on radius as $p\propto \varpi^{-5/2}$. 
Most of the self-similar solutions for thick discs such as the advective disc
of \citet{NY95} and slim-disc models at large mass accretion rates
\citep{slim} reproduce this dependence. Thus, thick barotropic locally-viscous
discs should have equatorial outflows. 

\subsection{General case of isotropic viscosity}

A more comprehensive and realistic assumption is that the angular momentum
flux is proportional to the local angular velocity gradient, and to use the
vertical rotation profile derived in Section~\ref{sec:rot}. Radial dependence
of angular frequency is assumed Keplerian $\Omega \propto \varpi^{-3/2}$, as
deviations from the Kepler's law are of the order
$\left(H/\varpi\right)^2$. We will also use the effective polytropic scaling
proposed in Section~\ref{sec:aansatz}. Vertical viscosity profile will be
described through the free parameter $t$, $\nu \propto (1-\xi^2)^t$.  For angular momentum flux, we can
rewrite equation~(\ref{E:wik}) as
\begin{equation}\label{E:ozwik}
\displaystyle e^\varphi_i \nabla_k {\rm {\rm w}_{ik}} = \left(-\frac{3}{2} \left( \Gamma_\nu + \Gamma_\rho + \frac{1}{2} + 2\Gamma_H (a+t) \frac{\xi^2}{1-\xi^2}\right)+ \frac{a\Gamma_p -(a+1)\Gamma_\rho}{2(a+1)}\left(1-2(a+t)\frac{\xi^2}{1-\xi^2}\right)\right)\frac{\nu\rho\Omega}{\varpi}.
\end{equation}
Here, additional terms proportional to $\xi^2/(1-\xi^2)$ arise from $\nu\rho$ dependence on height. 
Consequently, for the radial velocity structure, (\ref{E:angmo}) becomes
\begin{equation}\label{E:vrest}
\displaystyle v_\varpi = 2\frac{\nu}{\varpi} \left(A+C\frac{\xi^2}{1-\xi^2}\right),
\end{equation}
where 
\begin{equation}\label{E:acoeff}
A=\frac{a\Gamma_p -(a+1)\Gamma_\rho}{2(a+1)} - \frac{3}{2}\left( \Gamma_\nu + \Gamma_\rho + \frac{1}{2}\right)
\end{equation}
and
\begin{equation}\label{E:ccoeff}
C=-(a+t)\left( 3\Gamma_H+\frac{a\Gamma_p -(a+1)\Gamma_\rho}{a+1}\right).
\end{equation}
Coefficient $A$ has the physical meaning of normalized midplane velocity. The
second term in~(\ref{E:vrest}) does not diverge at the surface of the disc if
$t\geq 1$, because the viscosity itself $\nu \propto (1-\xi^2)^t$. Depending
on the quantity of $t$, velocity may diverge at the surface, but the matter
and angular momentum flux are always finite if $a+t\geq 0$. For different
model setups, we give the values of $A$ and $C$ in table~\ref{tab:varpars}.

{

  The standard-disc scaling for vertically-integrated viscous stress,
  $\displaystyle \int \rho \nu \varpi \pardir{\varpi}{\Omega} dz \propto \int p dz$, may be used to simplify the
  expressions. In our formalism, it is equivalent to
  \begin{equation}\label{E:SDgamma}
    \Gamma_\rho + \Gamma_\nu -\frac{3}{2} = \Gamma_p.
  \end{equation}
For the case of a steady-state disc this may be supplemented with the
condition for vertically-integrated viscous stress ignoring the corrections
due to the disc inner boundary, $\displaystyle W_{r\varphi} =
\frac{1}{2\pi}\dot{M}\Omega \propto R^{-3/2}$. For the thickness determined by
the vertical balance, $\Gamma_H = \frac{1}{2}\left( \Gamma_p - \Gamma_\rho +
3\right)$, and the standard-disc scaling for the viscous stress, this gives
  \begin{equation}\label{E:stat:gammarho}
    \Gamma_\rho^{\mbox{stationary}} = 3\Gamma_p+6,
  \end{equation}  
   \begin{equation}\label{E:stat:gammanu}
    \Gamma_\nu^{\mbox{stationary}} = -2\Gamma_p-\frac{9}{2},
  \end{equation}
   and thus
    \begin{equation}\label{E:stat:A}
      A^{\mbox{stationary}} = \frac{a}{2(a+1)} \Gamma_p - 3\left( \Gamma_p +2\right).
  \end{equation}
  
\subsection{Condition for an outflow}

It is straightforward to check that the expression for the midplane velocity,
\begin{equation}\label{E:vrest}
  v_\varpi(z=0)=2A\nu/\varpi,
\end{equation}
coincides with expression (6) from \citet{rafikov}. It is interesting to
derive a physically motivated criterion for a midplane outflow. Midplane
velocity is positive when
\begin{equation}\label{E:apos}
\displaystyle \Gamma_p \frac{a}{a+1} > 4\Gamma_\rho +3\Gamma_\nu +\frac{3}{2}.
\end{equation}
We can safely assume pressure decreasing outwards ($\Gamma_p<0$) and restrict
the possible solutions to positive $a$, as density is expected to decrease
with height. Let us also restrict the possible solutions with the stationary
standard disc scaling (\ref{E:stat:A}), containing only one free parameter
$\Gamma_p$. Condition (\ref{E:apos}) becomes
\begin{equation}\label{E:apos:std}
\displaystyle \Gamma_p \frac{a}{a+1} > 6\left( \Gamma_p + 2 \right).
\end{equation}
Evidently, an outflow does not exist for any positive value of $a$ if
$\Gamma_p > -2$. If $\displaystyle \Gamma_p < -\frac{12}{5} = -2.4$, a
midplane outflow exists for all the possible $a$. Between these values, an
outflow exists if
\begin{equation}\label{E:acrit}
a > -\frac{1+\Gamma_p/2}{1+\Gamma_p/2.4}.
\end{equation}
Vertical structure parameter affects the outflow condition only slightly, the
main constraints are set upon the radial profiles of certain variables, most
importantly -- on the viscous stress, or pressure, if the alpha-disc scaling
works. 

The left-hand side of equation~(\ref{E:apos:std}) may be also re-written using entropy
defined as $\displaystyle s=\ln \frac{p}{\rho^\gamma}$,  
\begin{equation}\label{E:apos:entropy}
\displaystyle \frac{\Gamma_p}{\gamma}\left( 1+s^{\prime\prime}\right)  > 6\left( \Gamma_p + 2 \right),
\end{equation}
where the second entropy derivative is normalized according to \citet{rafikov}
\begin{equation}
\displaystyle s^{\prime\prime}  > \frac{1}{2(a+1)} \pardirsec{\xi}{s} = \frac{1}{2(a+1)}
\pardirsec{\xi}{}\ln\left(\frac{p}{\rho^\gamma}\right)
\end{equation}
and the choice of
adiabatic index $\gamma$ is arbitrary as it enters only the definition of
$s$. Solving for the second derivative of entropy yields
\begin{equation}\label{E:apos:entropy}
\displaystyle s^{\prime\prime} > 6\gamma-1 + \frac{12\gamma}{\Gamma_p},
\end{equation}
that coincides with the condition (22) from \citet{rafikov} in the particular
case of stress proportional to pressure. However, the
possible values of $s^{\prime\prime}$ are limited. The possible effective
equations of state are limited by constant density with height, on one hand
($a>0$, $s^{\prime\prime}>-1$), and by the isothermal effective equation of state, on the other ($a<
+\infty$, $s^{\prime\prime}<\gamma-1$). Some of these solutions, with
$s^{\prime\prime}<0$, are convectively unstable, but still
possible because vertical convection is inefficient \citep{SSZ}. 
}

\subsection{The role of vertical structure}

Of all the possible values of $a$ and $t$, two cases are the most interesting: the
local-alpha case and the constant-stress case. First is important as the
simplest generalization of alpha assumption: viscous stress ${\rm
  w}_{r\varphi} = \frac{3}{2}\nu \rho \Omega$ is assumed proportional to local
pressure that is equivalent to $\Gamma_\nu + \Gamma_{\rho} -3/2 = \Gamma_p$
(that reproduces radial behaviour, see above equation~\ref{E:SDgamma})
and $t=1$ (vertical profile). The radial
scaling for $\Gamma_\nu$ should be also applicable in a more general case,
wherever the standard-disc assumption for vertically-integrated quantities
${\rm W}_{r\varphi} \propto \Pi$ holds.

The other possibility, vertically constant viscous stress, is reproduced by
assuming $a+t=0$ (as the stress is $\sim \nu \rho \Omega$). 
Physically it is better motivated than local-alpha. Magnetic
fields produced by non-linear MRI are expected to be relatively independent of
$z$ and easily transported between different $z$. Large-scale vertical
magnetic field should retain its strength to fulfill the flux conservation
rule, that suggests constant with height magnetic stresses created by
large-scale fields. Vertical structure for a
magnetized disc with magnetic field amplitudes independent of vertical
coordinate was considered by \citet{uzdensky13}. 
Models with $a+t\leq 0$, however, can not reproduce correctly the dynamics of
a standard disc, as there is an angular momentum flux through the surface. A
compromise we will use in this paper where we restrict ourselves to
conservative solutions is to consider the right limit $a + t \to +0$. 

As in the barotropic case, radial velocity in this case does not depend on
$z$, 
\begin{equation}\label{E:vrest:const}
\displaystyle v_\varpi^{\mbox{constant stress}} = 2\left(\frac{a\Gamma_p -(a+1)\Gamma_\rho}{2(a+1)} - \frac{3}{2}\left(\Gamma_p + 2\right)\right) \frac{\nu}{\varpi}.
\end{equation}
Depending on the radial structure of the disc and on the value of $a$, the
value of $v_\varpi$ and its sign may be different. In table~\ref{tab:varpars},
we consider several cases important for disc accretion. Along with the
standard-disc zones A, B, and C (see \citealt{SS73}) with different
assumptions about the vertical structure in density and viscosity, we apply
our calculations for the self-similar scalings of a thick quasi-spherical disc
and find that, for a broad range of possible parameters, midplane velocity and
pressure-weighed radial velocity (the importance of which will be shown in
Section~\ref{sec:heat}) are mostly positive. This is in contrast with the
classical zone A where the radial velocity is always negative. In
gas-pressure-dominated discs, radial velocity sometimes changes sign with
vertical coordinate, sometimes not. Except radiation-pressure-dominated thin
disc case, midplane velocity is always larger (either positive or smaller by
the absolute value) then average. 

\begin{table}\centering
\caption{Vertical structure for parameters relevant for different accretion
  disc models. ``SD'' refers to the standard disc model of
  \citet{SS73}. Averaged velocities are given with accuracy of about one per
  cent of $\nu_{\rm c}/H$. Different values of $a$ and $t$ are considered,
  $t=-a$ reproducing the constant-stress case (we use the $t+a \to +0$ limit,
  as described in the text), and $t=1$ for the local-alpha approximation. }\label{tab:varpars}
\begin{tabular}{lccccccccccc}
\hline
& $\Gamma_\rho$ & $\Gamma_p$ & $\Gamma_H$ & $a$ & $t$ & $a\Gamma_p -(a+1)\Gamma_\rho$ & $\Gamma_\nu$ & $A$ & $C$  & $\langle v_\varpi\rangle_\rho H/\nu_{\rm c}$ & $\langle v_\varpi\rangle_p H/\nu_{\rm c}$ \\
\hline
SD zone A & $3/2$ & $-3/2$ & $0$ & $3$ & $-3$ & $-21/2$ & $-3/2$ & $-33/16$ &
$0$  & $-3.28$ & $-6.77$ \\
 & $3/2$ & $-3/2$ & $0$ & $1$ & $1$ & $-9/2$ & $-3/2$ & $-15/8$ & $0$  & $-1.20$ & $-1.93$ \\
\\
SD zone B & $-33/20$ & $-51/20$ & $21/20$ & $3$ & $-3$ & $-21/20$ & $3/5$ & $111/160$ & $0$  & $-3.28$ & $2.27$ \\
 & $-33/20$ & $-51/20$ & $21/20$ & $3$ & $1$ & $-21/20$ & $3/5$ & $111/160$ & $-231/20$  & $-1.33$ & $0.84$ \\
\\
SD zone C & $-15/8$ & $-21/8$ & $9/8$ & $3$ & $-3$ & $-3/8$ & $3/4$ & $57/64$ & $0$  & $-3.28$ & $2.92$ \\
& $-15/8$ & $-21/8$ & $9/8$ & $3$ & $1$ & $-3/8$ & $3/4$ & $57/64$ & $-105/8$  & $-1.33$ & $-0.77$ \\
\\
quasi-spherical & $-3/2$ & $-5/2$ & $1$ & $0$ & $0$ & $3/2$ & $1/2$ & $3/2$ & $0$  & $-1.50$ & $3.0$ \\
& $-3/2$ & $-5/2$ & $1$ & $0$ & $1$ & $3/2$ & $1/2$ & $3/2$ & $-9/2$  & $-1.00$ & $0.60$ \\
 & $-3/2$ & $-5/2$ & $1$ & $3$ & $-3$ & $-3/2$ & $1/2$ & $9/16$ & $0$  & $-3.28$ & $1.85$ \\
 & $-3/2$ & $-5/2$ & $1$ & $3$ & $1$ & $-3/2$ & $1/2$ & $9/16$ & $-21/2$  & $-1.33$ & $-0.88$ \\
 \\
 convective & $-1/2$ & $-3/2$  &  $1$  &  $5/2$  &  $-5/2$ &  $-2$ &  $1/2$ & $-29/28$ & $0$ & $-9.17$  &  $-3.22$ \\ 
  & $-1/2$ & $-3/2$  &  $1$  &  $5/2$  &  $1$ &  $-2$ &  $1/2$ & $-29/28$ & $-17/2$ & $-3.94$  &  $-3.56$ \\ 
  & $-1/2$ & $-3/2$  &  $1$  &  $2$  &  $-2$ &  $-3/2$ &  $1/2$ & $-1$ & $0$ & $-8.43$  &  $-2.92$ \\ 
  & $-1/2$ & $-3/2$  &  $1$  &  $2$  &  $1$ &  $-3/2$ &  $1/2$ & $-1$ & $-15/2$ & $-3.86$  &  $-3.44$ \\ 
\hline
\end{tabular}
\end{table}

Apart from standard thin disc approximations, we estimate the properties of
the two disc models that in fact assume $H\sim R$. One of these scalings arise
if one assumes radial velocity scaling with $\Omega \varpi$ as in
advection-dominated solutions (in table~\ref{tab:varpars}, we refer to this
regime as to quasi-spherical). The other arises from the assumption of strong
radial convection. 
Convection-dominated discs \citep{QG00} tend to have $\Gamma_p \sim -3/2$,
hence a midplane outflow is not expected. However, the convection-dominated
solution assumes exact gyrentropic rotation throughout the volume of the
disc. Such an equilibrium requires strong convective motions, and neglects
other energy transfer processes that can affect the vertical and radial
structure of the model. Besides, convective models assume that the velocity
field can be easily separated into steady-state smoothly varying part and a
fluctuating part providing convective transport. As some simulations of thick
discs with convection show \citep{ECK88,igu02}, large convective cells may
contribute to velocity field even after averaging in time and radial
coordinate. Simulation results also suggest that convection is not efficient
enough to lead to gyrentropic rotation. 

\section{Radial heat transport}\label{sec:heat}

\subsection{Effective advection velocity}

If energy transfer is dominated by diffusive radiation transfer, and adiabatic
heating is neglected, energy equation may be written as follows
\begin{equation}\label{E:eneq}
\displaystyle \pardir{t}{\varepsilon} + \nabla \vector{F} = q,
\end{equation}
where $q$ is energy dissipation per unit volume, $\varepsilon$ is radiation
energy density, and $\displaystyle \vector{F}=-D\nabla \varepsilon +
\vector{v}\varepsilon$ is radiation energy flux consisting of a diffusion and
an advection parts. In the steady state, $\displaystyle
\pardir{t}{\varepsilon}$ disappears. To measure the effects of radial
advection and diffusion, we should integrate equation (\ref{E:eneq}) over $z$
taking into account the flux $Q_{\rm rad}$ radiated from the surface
\begin{equation}\label{E:QQ}
\displaystyle \ppardir{\varpi}{\int D \pardir{\varpi}{\varepsilon} dz - \int \varepsilon v_\varpi dz} = Q_{\rm rad}-Q_{\rm tot} ,
\end{equation}
where $Q_{\rm tot} = \int q dz$. 
This expression estimates the difference between the fluxes generated inside
the accretion disc and radiated from its surface. First term in the left-hand
side is suppressed when $\tau\gg 1$, but the second, advective, is potentially
important inside the disc. The advective term may be written as $Q_{\rm adv} =
\varepsilon \langle v_\varpi\rangle_\varepsilon$, where the velocity is
averaged with the weights proportional to the energy density
\begin{equation}\label{E:vaver:p}
\displaystyle \langle v_\varpi\rangle_\varepsilon = \frac{\int \varepsilon v_\varpi dz }{\int \varepsilon dz}.
\end{equation}
Evidently, averaging with $\varepsilon$ (or radiation pressure $p_{\rm rad} =
\varepsilon/3$) does not in general produce the same result as averaging over
$\rho$. As we should reproduce the standard disc equations after integrating
the dynamical equations over $z$,
\begin{equation}\label{E:vaver:rho}
\displaystyle \langle v_\varpi\rangle_\rho = \frac{\int \rho v_\varpi dz }{\int \rho dz}
\end{equation}
should be always negative, at least when $a+t>0$ and there is no stress at the disc surface. 

These suggestions are confirmed by direct integration in effective-polytrope approximation. For the velocity expressed by equation~(\ref{E:vrest}),
\begin{equation}
\displaystyle \frac{\dot M}{2\pi \varpi} = \Sigma \langle v_\varpi\rangle_\rho = \int \rho v_\varpi dz = \frac{2}{\varpi} \int \rho \nu \left( A+ C \frac{\xi^2}{1-\xi^2}\right) dz = \frac{2}{\varpi} \rho_{\rm c} \nu_{\rm c} H \left( \left(A-C\right) \gfun{a+t} + C \gfun{a+t-1}\right),
\end{equation}
where 
\begin{equation}
\displaystyle \gfun{n} = \int_{-1}^{1} (1-\xi^2)^n d\xi =  \frac{\sqrt{\pi}\Gamma\,(n+1)}{\Gamma\,(n+3/2)}.
\end{equation}
Substituting $A$ from~(\ref{E:acoeff}) and $C$ from~(\ref{E:ccoeff}), 
\begin{equation}\label{E:mdot}
\displaystyle 
\Sigma \langle v_\varpi\rangle_\rho = -\frac{3}{\varpi} \rho_{\rm c} \nu_{\rm c} \gfun{a+t} \left( \Gamma_\nu +\Gamma_\rho + \Gamma_{\rm H} + \frac{1}{2}\right).
\end{equation}
The last bracket is proportional to $\ppardir{\varpi}{\nu\rho H \Omega
  \varpi^2} \propto \ppardir{\varpi}{{\rm W}_{r\varphi} \varpi^2}$ as it
should be in standard theory, and thus does not change sign in a stationary
case.

The case of constant viscous stress should be treated separately. Equation~(\ref{E:mdot}) can not be reproduced by simply setting $t=-a$, because in this case angular momentum transfer is not conservative. However, it is possible to consider the above set of formulae in the limit $a+t\to +0$ which produces a different result. Indeed, it is easy to show that
\begin{equation}
\lim_{n\to +0} n \gfun{n-1}=1, 
\end{equation}
hence, for the constant-stress case, equation~(\ref{E:vrest}) becomes
\begin{equation}
\displaystyle 
\langle v_{\varpi}\rangle_{\rho \mbox{, constant stress}}= \frac{2}{\varpi} \rho \nu \left( A+\frac{C}{2} \right),
\end{equation}
that is easily reduced to equation~(\ref{E:mdot}) if the values of $A$ and $C$
are substituted using equations~(\ref{E:acoeff}) and (\ref{E:ccoeff}). In this limit, the velocity field becomes unphysical: velocity is constant and, generally, different from the mean velocity value. 
In particular, radial velocity may be positive in every internal point, while
the mass transfer is dominated by the infinitely thin surface layer moving
inwards. It should be noted that, if in real accretion discs viscous stresses
are constant with height, there should be a non-negligible surface stress and
thus angular momentum transfer to wind or corona. In our formalism this
corresponds to direct substitution of $t=-a$ to equation~(\ref{E:ozwik})
before integration instead of approaching the limit as it was made above. In
this case, $\displaystyle v_\varpi = \frac{2A}{\varpi} \frac{\rho
  \nu}{\Sigma}$. If $A<0$, this is easily interpreted as angular momentum loss
from the disc surface, while the case of $A\geq 0$ does not have any evident
physical meaning as there should not be any angular momentum source at the
surface of the disc. 

For radiation-pressure-dominated regime, we can replace radiation energy density with pressure and calculate $\langle v\rangle_{\varepsilon}$ as
\begin{equation}
\begin{array}{l}
\displaystyle \langle v_\varpi\rangle_{\varepsilon} = \frac{\int p v_\varpi dz}{\int p dz} = \frac{\nu_{\rm c}}{ H \gfun{a+1}}\left( (A-C) \gfun{a+t+1} + C \gfun{a+t}\right)  \\
\displaystyle \qquad{} =\frac{1}{a+t+\frac{3}{2}}\frac{\nu_{\rm c}\gfun{a+t}}{ H \gfun{a+1}}\left( -\frac{3}{2}  \left(a+t+1\right) \left(\Gamma_\nu + \Gamma_\rho +\frac{a+t}{a+t+1}\Gamma_H + \frac{1}{2}\right) + \frac{a\Gamma_p-(a+1)\Gamma_\rho}{2(a+1)} \right).\\
\end{array}
\end{equation}
For the particular case of ``local-alpha'' viscosity when local stress ${\rm w}_{r\varphi} \propto p$, $t=1$, and the radial gradients conforming to the quasi-spherical case $\Gamma_\rho = -3/2$, $\Gamma_{H}=1$, $\Gamma_p = -5/2$, $\Gamma_\nu = 1/2$:
\begin{equation}
\displaystyle \langle v\rangle_{\varepsilon, \mbox{ quasi-spherical, local-alpha}} = -\frac{3}{2} \frac{a^2-\frac{5}{3}a-1}{(2a+5)(a+1)} \frac{\nu_{\rm c}}{H},
\end{equation}
that changes sign at $a=(\sqrt{61}-5)/6\simeq 0.47$, the stiffer effective
polytrope corresponding to positive weighed velocity. Hence, pure
``local-alpha'' condition easily reproduces inward net advection, always but
for exotic, very stiff vertical structure. However, in the more realistic
assumption of vertically-constant viscous stress ($t=-a$),
\begin{equation}
\displaystyle \langle v\rangle_{\varepsilon, \mbox{ quasi-spherical, constant stress}} = \frac{1}{3} \frac{a+6}{(a+1)\gfun{a+1}} \frac{\nu_{\rm c}}{H},
\end{equation}
that is always positive for positive $a$. 

In Fig.~\ref{fig:verstre:ab}, we show not only the density profiles, but also
angular velocities and radial velocities calculated using
equations~(\ref{E:app:omega}) and (\ref{E:app:vr}), respectively. Effective
polytropic approximation with $a=3$ accurately predicts the profile of
$\Omega$, as well as the midplane value of $v_\varpi$, though the velocities
at larger heights may deviate strongly due to differences in density profile. 

The estimates made above should also be valid for optically thin, radiatively
inefficient discs, if thermal conduction is ignored.
Thermal conduction will level out temperature gradients, and the likely
vertical structure will have $a\gg 1$ (with the case $a\to \infty$
corresponding to constant temperature). Large $a$ in a quasi-spherical disc
means negative or positive $\displaystyle \langle v\rangle_{\varepsilon}$,
depending on the energy release dependence on height. In all the cases,
pressure-averaged radial velocity is larger than density-averaged.

\subsection{Observational implications}\label{sec:obs}

Consequences of non-trivial vertical structure may be well understood in the
standard approach to advective discs used, for instance, in
\citet{lipunova}. The difference between the radiation flux generated inside
the disc and that radiated from its surface is expressed through entropy
gradient as
\begin{equation}\label{E:qadv}
Q_{\rm adv} = \int v_\varpi \rho T \pardir{r}{s} dz,
\end{equation}
or, for radiation-pressure-dominated medium,
\begin{equation}\label{E:qadv:rad}
Q_{\rm adv} = \int v_\varpi \rho \left(\ppardir{r}{\frac{\varepsilon}{\rho}} +
\frac{1}{3} \varepsilon \ppardir{r}{\frac{1}{\rho}} \right) dz.
\end{equation}
If the variables are separable, the result is
\begin{equation}\label{E:qadv:separad}
Q_{\rm adv} = \gfun{a+1} \langle v_\varpi \rangle_\varepsilon  \Sigma\left(
\ppardir{\varpi}{\frac{3\Pi}{\Sigma}} + \Pi
\ppardir{\varpi}{\frac{1}{\Sigma}} + \frac{\Pi}{\Sigma} \pardir{\varpi}{\ln H}\right).
\end{equation}
Standard way to use this expression is to replace the mean radial velocity
with the density-averaged value (that also absorbs surface density). We
can estimate the effects of vertical structure by introducing a discrepancy
factor 
\begin{equation}\label{E:betafac}
\beta = \frac{\langle v_\varpi \rangle_\varepsilon}{\langle v_\varpi \rangle_\rho},
\end{equation}
and assuming power-law dependences on radius for $\Pi$ and $\Sigma$. Thus, 
\begin{equation}\label{E:qadv:betafac}
Q_{\rm adv} = -\gfun{a+1} \beta \left( 3\Gamma_\Pi - 4\Gamma_\Sigma+\Gamma_H\right) \frac{\dot{M}}{4\pi \varpi} \frac{\Pi}{\varpi\Sigma}. 
\end{equation}
The simplest observational manifestation arises from energy balance, $Q_{\rm
  tot} = Q_{\rm adv} + Q_{\rm rad}$. 
The observed effective temperature distribution will change affected by radial
heat transport as
\begin{equation}\label{E:qadv:teff}
\sigma_{\rm B} T_{\rm eff}^4 \simeq \frac{3}{8\pi} \frac{GM\dot{M}}{\varpi^3}
\left( 1 - \frac{16}{15}\beta \frac{\varpi}{GM} \frac{\Pi}{\Sigma}\right).
\end{equation}
This is a rough estimate based on the supposedly power-law radial dependences
of all the quantities (we assumed $a=1$, $\Gamma_\Sigma = -1/2$, $\Gamma_\Pi =
-3/2$). It does however show clearly that the corrections that advection
introduces in the temperature profile of the disc depend on the ratio of 
average velocities, and change sign when $\langle v_\varpi
\rangle_\varepsilon$ becomes directed outwards. { The bracket in
  expression~(\ref{E:qadv:teff}) also estimates the overall change in
  accretion efficiency. }


{
  Change in efficiency is essential for geometrically-thick,
  advection-dominated flows that should generally conform to the
  quasi-spherical scalings discussed above.
  Such flows are expected to form
when cooling is not efficient enough and large part of the released heat is
stored inside the disc. Sound speed is limited only by the virial velocity,
that implies $H/\varpi \sim 1$, and also suggests virial scaling for other
velocity
components, and hence $\Gamma_\rho =-3/2$ and $\Gamma_p =-5/2$.
As temperatures and radial velocities are high, advection is very important
for any type of quasi-spherical solutions. Such flows are expected
to appear in two cases: very high and very low mass accretion rates. High mass
accretion rate leads to a high luminosity and optical depth, and the photon
diffusion time becomes comparable to the viscous timescale. This is the case
of slim discs \citep{abram88, slim} and other solutions proposed for
super-Eddington accretion \citep{poutanen07}. Advection effects are known to
limit the luminosity by storing the heat inside the flow absorbed by the black
hole. Reverse advection, however, smoothens this effect by spreading the
trapped photons to larger disc radii and thus increasing the total
luminosity. Radiative efficiency becomes close to that of a standard disc. 

The other case is advection-dominated flows expected to
exist at low mass accretion rates, about $10^{-2}$ of Eddington and lower
\citep{YN_review}. Here, transporting the hot gas outwards can again increase the
overall efficiency of the flow as the cooling timescale becomes shorter with
respect to the viscous and dynamical times at larger radii \citep{NY95more}.
In both cases we expect the flow efficiency to increase due to lower rate of heat
advection. However, any type of a radiatively inefficient quasi-spherical flow
also differs from the thin disc in sense of dynamics, that we will discuss
later in section~\ref{sec:disc:thick}. 
}

\section{Discussion}\label{sec:disc}

\subsection{Anisotropic and non-local viscosity}\label{sec:disc:aniso}

One principal assumption of this work is that viscosity is of Newtonian type,
with the stress proportional to strain with equal proportionality coefficients
in different directions.
This case may be also described as isotropic viscosity, because isotropic
turbulent motions reproduce this scaling, and the fourth-rank viscosity tensor
reduces to a scalar.
However, there are at least two cases when angular momentum transfer can be
anisotropic: anisotropic turbulence (created, for instance, by convection) and
large-scale magnetic fields. Small-scale chaotic magnetic fields can also
produce anisotropic viscosity if their statistical properties favour certain
directions in space.

On one hand, anisotropic turbulence can easily change the direction of
angular momentum transfer (as suggested by \citealt{wasiutynski}; for a very
general but purely hydrodynamical consideration, see~\citealt{tassoul},
section 8.5). 
On the other, chaotic magnetic fields produce viscous stresses
proportional to strain components. Following \cite{ogilvie03}, off-diagonal
magnetic stress components may be written in linear approximation as
\begin{equation}
{\rm w}_{(\varpi,z)\varphi} \propto - \varpi B_{\varpi,z} B_\varphi \pardir{(\varpi,z)}{\Omega }.
\end{equation}
Stress components are thus proportional to the corresponding strains, and all the conclusions about the radial angular momentum transfer hold. However, the vertical term must in general be rescaled with the corresponding field component. Numerical simulations of MHD turbulence in saturated state suggest $B_{\varpi,z} \ll B_{\varphi}$ in Keplerian shear flows \citep{PB13}. Similarly, turbulent velocity dispersion is expected to be dominated by azimuthal motions \citep{ogilvie03}. For an anisotropic turbulent velocity field, $\varpi \varphi$-stress is proportional to 
\begin{equation}
\displaystyle {\rm w}_{\varpi \varphi} \propto \frac{1}{\varpi}\left\langle v_{\mbox{turbulent}, r}^2\right\rangle \ppardir{\varpi}{\varpi^2 \Omega} - 2\left\langle v_{\mbox{turbulent}, \varphi}^2\right\rangle \Omega.
\end{equation}
Angular momentum transfer direction for predominantly azimuthal velocity field has thus the same direction as for the isotropic case, but becomes proportional to $\Omega$ rather than its derivative. 

Angular momentum transfer becomes non-local if the spatial scales of the
magnetic loops and mixing lengths for turbulent motions become comparable to
or larger than disc thickness.
{ Extremely non-local angular momentum transfer is provided by
  non-axisymmetric global perturbation modes in the disc. In particular,
  spiral shock waves were recently shown \citep{jiang17} to play important role in
  radiation-pressure-dominated accretion discs. By definition, such
  structures are non-linear, non-axisymmetric and non-stationary. This
  means that the velocity averaging procedure used in section~\ref{sec:heat} becomes
  invalid, as there are local heat flows and non-stationary processes
  connected non-linearly with the local perturbations. More generally,
  existence of large-scale structures in the disc favoured by many simulations
  \citep{jiang14,sana16} will lead to additional correction factors for the averaged
  velocity values, most likely increasing the heat transport outwards when the
  radial entropy gradient is super-adiabatic. For non-local viscous stresses
  and large-scale inhomogeneities in the disc, it is still possible to
  describe the angular momentum transfer in terms of Reynolds stress, but its
  components should depend on the conditions throughout the disc, not only at
  the radius considered. The $r\varphi$ component of the Reynolds stress is
  expected to scale approximately with pressure \cite[section~3.3]{balbus03} that means
  one should expect ``local-alpha'' to be a reasonable description for the
  average circulation pattern. However, strong deviations from linearity,
  stationarity and axisymmetry may alter the results. 
}

In the limiting case of large-scale dynamically
important magnetic fields and turbulent motions in thick discs, ${\rm
  w}_{\varpi z}$ stress component should be
also important. In particular, ${\rm w}_{\varpi z}$ is likely responsible for the
smooth radial velocity profile, very weakly dependent on $z$, in some of the
MHD simulations \citep{sadowski14,sadowski16,jiang14}. 
At the same time, certain simulations of thin ($H/R \sim 0.1$) MHD accretion
discs (such as \citealt{zhustone}) tend to reproduce the vertical profiles of
radial velocities. Possibly, some disc thickness effects smooth out the radial
velocity profiles. 





\subsection{Effects of disc thickness}\label{sec:disc:thick}

Directly, effects of disc thickness are not expected to affect much the
results given in this paper. The equation for angular momentum transfer
(equation~\ref{E:wik}) as well as the vertical derivative of angular frequency
estimate (equation~\ref{E:oz}) are valid in general case. All the results near
the equatorial plane hold, though the midplane value of angular frequency may
change significantly. In particular, the statement about the midplane outflow
does not rely strongly on the assumption about disc thickness as all the
neglected terms are proportional to $\left(z/\varpi\right)^2$ rather than to
$\left(H/\varpi\right)^2$.

A potentially important issue is the poloidal velocity field. So far we neglected the
vertical motions in the disc. They should be negligible near the
midplane, but at $z\sim H$ the impact of $v_z$ is
important. Equation~(\ref{E:angmo}) becomes
\begin{equation}\label{E:zangmo}
\frac{1}{\varpi}\rho v_\varpi \ppardir{\varpi}{\Omega \varpi^2} + \varpi \rho v_z \pardir{z}{\Omega}= e^\varphi_i \nabla_k {\rm {\rm w}_{ik}}.
\end{equation}
For a conservative disc, vertical velocities should follow the changes in disc
thickness, approximately as $\displaystyle v_z = \Gamma_H \frac{z}{\varpi}
v_\varpi$. This makes a correction to the expression for steady-state radial
velocity {
\begin{equation}\label{E:vradthick}
\displaystyle v_\varpi =
v_\varpi^{\mbox{thin-disc}}\cdot\left(1+\frac{(a\Gamma_p-(a+1)\Gamma_\rho)\Gamma_H}{a+1}
\left(\frac{z}{\varpi}\right)^2\right).
\end{equation}
This correction multiplier is unlikely to alter the overall circulation
pattern, especially if $\frac{(a\Gamma_p-(a+1)\Gamma_\rho)\Gamma_H}{a+1}
>0$, that is true for most of the models considered. We suggest that
the effects of $\varpi z$ viscosity component, angular momentum exchange with
the wind or corona, and non-local angular momentum
transfer are more important.
}

\section{Conclusions}

We conclude that, in a thin accretion disc with local, isotropic viscosity and no angular momentum loss from the surface, poloidal velocity field is strongly affected by meridional circulation. The principal quantity is the radial slope of the midplane viscous stress ${\rm w}_{r\varphi}$. If the stress decreases with radius as ${\rm w}_{r\varphi}\propto \varpi^{-2}$ or faster, a midplane outflow is possible. 

Existence of a midplane outflow in geometrically thick accretion discs should be important for advection effects, especially for very high (super-Eddington) mass accretion rates, when optical depth is large and temperature grows rapidly toward the equatorial plane. In thick discs one should expect the viscous stress to depend on the radial coordinate as ${\rm w}_{r\varphi}\propto \varpi^{-5/2}$ that evidently should lead to an outflow.
Depending on the details of vertical structure, energy dissipation and
transfer, advective heat flux may be dominated either by the overall radial
velocity directed inwards or by the equatorial counterflow. The effective
radial velocity of heat transfer may differ from the density-averaged velocity
by a factor of several.

\section*{Acknowledgments} 

I would like to thank Galina Lipunova, Alexander Philippov, Alexander
Tchekhovskoy, Phil Armitage, and Richard Nelson for valuable
discussions. { Special thanks to Vyacheslav Zhuravlev who has drawn my attention
to the issue of meridional circulation. The work was supported by the Academy
of Finland grant 268740 and the Russian Scientific Council grant 14-12-00146.}

\bibliographystyle{mnras}
\bibliography{mybib}

\hyphenation{Post-Script Sprin-ger}
\begin{thebibliography}{}
\makeatletter
\relax
\def\mn@urlcharsother{\let\do\@makeother \do\$\do\&\do\#\do\^\do\_\do\%\do\~}
\def\mn@doi{\begingroup\mn@urlcharsother \@ifnextchar [ {\mn@doi@}
  {\mn@doi@[]}}
\def\mn@doi@[#1]#2{\def\@tempa{#1}\ifx\@tempa\@empty \href
  {http://dx.doi.org/#2} {doi:#2}\else \href {http://dx.doi.org/#2} {#1}\fi
  \endgroup}
\def\mn@eprint#1#2{\mn@eprint@#1:#2::\@nil}
\def\mn@eprint@arXiv#1{\href {http://arxiv.org/abs/#1} {{\tt arXiv:#1}}}
\def\mn@eprint@dblp#1{\href {http://dblp.uni-trier.de/rec/bibtex/#1.xml}
  {dblp:#1}}
\def\mn@eprint@#1:#2:#3:#4\@nil{\def\@tempa {#1}\def\@tempb {#2}\def\@tempc
  {#3}\ifx \@tempc \@empty \let \@tempc \@tempb \let \@tempb \@tempa \fi \ifx
  \@tempb \@empty \def\@tempb {arXiv}\fi \@ifundefined
  {mn@eprint@\@tempb}{\@tempb:\@tempc}{\expandafter \expandafter \csname
  mn@eprint@\@tempb\endcsname \expandafter{\@tempc}}}

\bibitem[\protect\citeauthoryear{{Abramowicz}, {Czerny}, {Lasota}  \&
  {Szuszkiewicz}}{{Abramowicz} et~al.}{1988}]{abram88}
{Abramowicz} M.~A.,  {Czerny} B.,  {Lasota} J.~P.,   {Szuszkiewicz} E.,  1988,
  \mn@doi [\apj] {10.1086/166683}, \href
  {http://adsabs.harvard.edu/abs/1988ApJ...332..646A} {332, 646}

\bibitem[\protect\citeauthoryear{{Balbus}}{{Balbus}}{2003}]{balbus03}
{Balbus} S.~A.,  2003, \mn@doi [\araa]
  {10.1146/annurev.astro.41.081401.155207}, \href
  {http://adsabs.harvard.edu/abs/2003ARA%26A..41..555B} {41, 555}

\bibitem[\protect\citeauthoryear{{Eggum}, {Coroniti}  \& {Katz}}{{Eggum}
  et~al.}{1988}]{ECK88}
{Eggum} G.~E.,  {Coroniti} F.~V.,   {Katz} J.~I.,  1988, \mn@doi [\apj]
  {10.1086/166462}, \href {http://adsabs.harvard.edu/abs/1988ApJ...330..142E}
  {330, 142}

\bibitem[\protect\citeauthoryear{{Fromang}, {Lyra}  \& {Masset}}{{Fromang}
  et~al.}{2011}]{fromang11}
{Fromang} S.,  {Lyra} W.,   {Masset} F.,  2011, \mn@doi [\aap]
  {10.1051/0004-6361/201016068}, \href
  {http://adsabs.harvard.edu/abs/2011A%26A...534A.107F} {534, A107}

\bibitem[\protect\citeauthoryear{{Igumenshchev}}{{Igumenshchev}}{2002}]{igu02}
{Igumenshchev} I.~V.,  2002, \mn@doi [\apjl] {10.1086/344148}, \href
  {http://adsabs.harvard.edu/abs/2002ApJ...577L..31I} {577, L31}

\bibitem[\protect\citeauthoryear{{Jiang}, {Stone}  \& {Davis}}{{Jiang}
  et~al.}{2014}]{jiang14}
{Jiang} Y.-F.,  {Stone} J.~M.,   {Davis} S.~W.,  2014, \mn@doi [\apj]
  {10.1088/0004-637X/796/2/106}, \href
  {http://adsabs.harvard.edu/abs/2014ApJ...796..106J} {796, 106}

\bibitem[\protect\citeauthoryear{{Jiang}, {Stone}  \& {Davis}}{{Jiang}
  et~al.}{2017}]{jiang17}
{Jiang} Y.-F.,  {Stone} J.,   {Davis} S.~W.,  2017, preprint, \href
  {http://adsabs.harvard.edu/abs/2017arXiv170902845J} {} (\mn@eprint {arXiv}
  {1709.02845})

\bibitem[\protect\citeauthoryear{{Ketsaris} \& {Shakura}}{{Ketsaris} \&
  {Shakura}}{1998}]{KS98}
{Ketsaris} N.~A.,  {Shakura} N.~I.,  1998, \mn@doi [Astronomical and
  Astrophysical Transactions] {10.1080/10556799808201769}, \href
  {http://adsabs.harvard.edu/abs/1998A%26AT...15..193K} {15, 193}

\bibitem[\protect\citeauthoryear{{Lipunova}}{{Lipunova}}{1999}]{lipunova}
{Lipunova} G.~V.,  1999, Astronomy Letters, \href
  {http://adsabs.harvard.edu/abs/1999AstL...25..508L} {25, 508}

\bibitem[\protect\citeauthoryear{{Narayan} \& {Yi}}{{Narayan} \&
  {Yi}}{1995a}]{NY95}
{Narayan} R.,  {Yi} I.,  1995a, \mn@doi [\apj] {10.1086/175599}, \href
  {http://adsabs.harvard.edu/abs/1995ApJ...444..231N} {444, 231}

\bibitem[\protect\citeauthoryear{{Narayan} \& {Yi}}{{Narayan} \&
  {Yi}}{1995b}]{NY95more}
{Narayan} R.,  {Yi} I.,  1995b, \mn@doi [\apj] {10.1086/176343}, \href
  {http://adsabs.harvard.edu/abs/1995ApJ...452..710N} {452, 710}

\bibitem[\protect\citeauthoryear{{Ogilvie}}{{Ogilvie}}{2003}]{ogilvie03}
{Ogilvie} G.~I.,  2003, \mn@doi [\mnras] {10.1046/j.1365-8711.2003.06359.x},
  \href {http://adsabs.harvard.edu/abs/2003MNRAS.340..969O} {340, 969}

\bibitem[\protect\citeauthoryear{{Parkin} \& {Bicknell}}{{Parkin} \&
  {Bicknell}}{2013}]{PB13}
{Parkin} E.~R.,  {Bicknell} G.~V.,  2013, \mn@doi [\apj]
  {10.1088/0004-637X/763/2/99}, \href
  {http://adsabs.harvard.edu/abs/2013ApJ...763...99P} {763, 99}

\bibitem[\protect\citeauthoryear{{Philippov} \& {Rafikov}}{{Philippov} \&
  {Rafikov}}{2017}]{rafikov}
{Philippov} A.~A.,  {Rafikov} R.~R.,  2017, \mn@doi [\apj]
  {10.3847/1538-4357/aa60ca}, \href
  {http://adsabs.harvard.edu/abs/2017ApJ...837..101P} {837, 101}

\bibitem[\protect\citeauthoryear{Pitaevskii \& Lifshitz}{Pitaevskii \&
  Lifshitz}{2012}]{kinetics}
Pitaevskii L.,  Lifshitz E.,  2012, Physical Kinetics.
No. v. 10, Elsevier Science, \url
  {https://books.google.ru/books?id=DTHxPDfV0fQC}

\bibitem[\protect\citeauthoryear{{Poutanen}, {Lipunova}, {Fabrika}, {Butkevich}
   \& {Abolmasov}}{{Poutanen} et~al.}{2007}]{poutanen07}
{Poutanen} J.,  {Lipunova} G.,  {Fabrika} S.,  {Butkevich} A.~G.,   {Abolmasov}
  P.,  2007, \mn@doi [\mnras] {10.1111/j.1365-2966.2007.11668.x}, \href
  {http://adsabs.harvard.edu/abs/2007MNRAS.377.1187P} {377, 1187}

\bibitem[\protect\citeauthoryear{{Quataert} \& {Gruzinov}}{{Quataert} \&
  {Gruzinov}}{2000}]{QG00}
{Quataert} E.,  {Gruzinov} A.,  2000, \mn@doi [\apj] {10.1086/309267}, \href
  {http://adsabs.harvard.edu/abs/2000ApJ...539..809Q} {539, 809}

\bibitem[\protect\citeauthoryear{{S{\c a}dowski}}{{S{\c a}dowski}}{2011}]{slim}
{S{\c a}dowski} A.,  2011, ArXiv e-prints: 1108.0396, \href
  {http://adsabs.harvard.edu/abs/2011arXiv1108.0396S} {}

\bibitem[\protect\citeauthoryear{{S{\c a}dowski} \& {Narayan}}{{S{\c a}dowski}
  \& {Narayan}}{2016a}]{sana16}
{S{\c a}dowski} A.,  {Narayan} R.,  2016a, \mn@doi [\mnras]
  {10.1093/mnras/stv2941}, \href
  {http://adsabs.harvard.edu/abs/2016MNRAS.456.3929S} {456, 3929}

\bibitem[\protect\citeauthoryear{{S{\c a}dowski} \& {Narayan}}{{S{\c a}dowski}
  \& {Narayan}}{2016b}]{sadowski16}
{S{\c a}dowski} A.,  {Narayan} R.,  2016b, \mn@doi [\mnras]
  {10.1093/mnras/stv2941}, \href
  {http://adsabs.harvard.edu/abs/2016MNRAS.456.3929S} {456, 3929}

\bibitem[\protect\citeauthoryear{{S{\c a}dowski}, {Narayan}, {Tchekhovskoy},
  {Abarca}, {Zhu}  \& {McKinney}}{{S{\c a}dowski} et~al.}{2014}]{sadowski14}
{S{\c a}dowski} A.,  {Narayan} R.,  {Tchekhovskoy} A.,  {Abarca} D.,  {Zhu} Y.,
    {McKinney} J.~C.,  2014, preprint, \href
  {http://adsabs.harvard.edu/abs/2014arXiv1407.4421S} {} (\mn@eprint {arXiv}
  {1407.4421})

\bibitem[\protect\citeauthoryear{{Shakura} \& {Sunyaev}}{{Shakura} \&
  {Sunyaev}}{1973}]{SS73}
{Shakura} N.~I.,  {Sunyaev} R.~A.,  1973, \aap, \href
  {http://adsabs.harvard.edu/abs/1973A%26A....24..337S} {24, 337}

\bibitem[\protect\citeauthoryear{{Shakura}, {Sunyaev}  \&
  {Zilitinkevich}}{{Shakura} et~al.}{1978}]{SSZ}
{Shakura} N.~I.,  {Sunyaev} R.~A.,   {Zilitinkevich} S.~S.,  1978, \aap, \href
  {http://adsabs.harvard.edu/abs/1978A%26A....62..179S} {62, 179}

\bibitem[\protect\citeauthoryear{{Stoll} \& {Kley}}{{Stoll} \&
  {Kley}}{2016}]{stollkley}
{Stoll} M.~H.~R.,  {Kley} W.,  2016, \mn@doi [\aap]
  {10.1051/0004-6361/201527716}, \href
  {http://adsabs.harvard.edu/abs/2016A%26A...594A..57S} {594, A57}

\bibitem[\protect\citeauthoryear{Tassoul}{Tassoul}{2000}]{tassoul}
Tassoul J.,  2000, Stellar Rotation.
Cambridge Astrophysics, Cambridge University Press, \url
  {http://books.google.fi/books?id=yWh0MR\_xCg8C}

\bibitem[\protect\citeauthoryear{{Tayler}}{{Tayler}}{1973}]{tayler73}
{Tayler} R.~J.,  1973, \mn@doi [\mnras] {10.1093/mnras/165.1.39}, \href
  {http://adsabs.harvard.edu/abs/1973MNRAS.165...39T} {165, 39}

\bibitem[\protect\citeauthoryear{{Urpin}}{{Urpin}}{1984}]{urpin84}
{Urpin} V.~A.,  1984, \sovast, \href
  {http://adsabs.harvard.edu/abs/1984SvA....28...50U} {28, 50}

\bibitem[\protect\citeauthoryear{{Uzdensky}}{{Uzdensky}}{2013}]{uzdensky13}
{Uzdensky} D.~A.,  2013, \mn@doi [\apj] {10.1088/0004-637X/775/2/103}, \href
  {http://adsabs.harvard.edu/abs/2013ApJ...775..103U} {775, 103}

\bibitem[\protect\citeauthoryear{{Wasiutynski}}{{Wasiutynski}}{1946}]{wasiutynski}
{Wasiutynski} J.,  1946, Astrophysica Norvegica, \href
  {http://adsabs.harvard.edu/abs/1946ApNr....4....1W} {4, 1}

\bibitem[\protect\citeauthoryear{{Yuan} \& {Narayan}}{{Yuan} \&
  {Narayan}}{2014}]{YN_review}
{Yuan} F.,  {Narayan} R.,  2014, \mn@doi [\araa]
  {10.1146/annurev-astro-082812-141003}, \href
  {http://adsabs.harvard.edu/abs/2014ARA%26A..52..529Y} {52, 529}

\bibitem[\protect\citeauthoryear{{Zhu} \& {Stone}}{{Zhu} \&
  {Stone}}{2017}]{zhustone}
{Zhu} Z.,  {Stone} J.~M.,  2017, preprint, \href
  {http://adsabs.harvard.edu/abs/2017arXiv170104627Z} {} (\mn@eprint {arXiv}
  {1701.04627})

\makeatother
\end{thebibliography}

\appendix

\section{Vertical structure in constant-stress approximation}\label{sec:app:vert}

Here, we will consider the vertical structure for the particular case of
constant dissipation with height, $\nu \rho = const$. 
Basic equations are vertical hydrostatic equilibrium
\begin{equation}
\frac{dp}{dz} = - \Omega^2 \rho z,
\end{equation}
optical depth growth
\begin{equation}
\frac{d\tau}{dz} = - \varkappa \rho,
\end{equation}
where $\varkappa$ is (Rosseland) opacity, with contributions from free-free absorption and free electron scattering taken into account. Fick's law for radiation diffusion
\begin{equation}
F = - \frac{c}{\varkappa \rho} \frac{d}{dz}\left( p_{\rm r} \right)
\end{equation}
where $\displaystyle p_{\rm r} = \frac{4\sigma_{\rm SB}}{3c} T^4$ is radiation
pressure, $\sigma_{\rm SB}$ is Stefan-Boltzmann constant, and dissipation law
\begin{equation}\label{E:diss}
\frac{dF}{dz} =  \nu \rho \left( \left( \varpi
\pardir{\varpi}{\Omega}\right)^2 +\varpi^2 \left(\pardir{z}{\Omega}\right)^2\right)= \nu \rho \Omega^2 \left(
\frac{9}{4} + \left(\frac{a\Gamma_p - (a+1)\Gamma_\rho}{2(a+1)}\frac{z}{\varpi}\right)^2\right)\simeq \frac{9}{4} \nu \rho \Omega^2.
\end{equation}
Below we will neglect the $O(z/\varpi)^2$ term. 
Angular frequency is assumed indistinguishable from Keplerian and constant
with height (only deviations in second derivative $\pardirsec{z}{\Omega}$ are important in our assumptions). Together with constancy of dissipation, it allows to integrate equation~(\ref{E:diss}) directly as
\begin{equation}\label{E:dissint}
F = \frac{9}{4}\nu \rho \Omega^2 z.
\end{equation}

Density may be normalized as follows:
\begin{equation}
\rho = \frac{\Sigma}{H} \tilde\rho= \frac{2 \tau_0}{\varkappa_T H} \tilde\rho,
\end{equation}
where $\varkappa_{\rm T}\simeq 0.34{\rm cm^2 \,g^{-1}}$ is Thomson opacity, and $\tau_0=\varkappa_{\rm T} \Sigma /2 $
is the total Thomson opacity from the equatorial plane to the infinity along
the vertical coordinate. Analogously,
\begin{equation}
p = \frac{2\tau_0 H \Omega^2}{\varkappa_T} \tilde p. 
\end{equation}
The same normalization will be used for radiation pressure. 
The kinematic viscosity is determined by the largest scales and largest velocities reachable for MHD turbulence, hence $\nu \sim H c_{\rm s}$. It is convenient to use the same normalization as for pressure for the quantity
\begin{equation}
\nu \rho \Omega = \alpha  \frac{2\tau_0 H \Omega^2}{\varkappa_T},
\end{equation}
where $\alpha$ is a free parameter similar in physical meaning to the thin-disc $\alpha$ parameter. 

Energy radiated from one side of the disc may be used to link the mass
accretion rate with disc thickness as
\begin{equation}
\frac{3}{8\pi} \frac{GM\dot{M}}{R^3}= F(z=H) = \frac{9}{4} \nu \rho \Omega^2 H,
\end{equation}
where we neglected the correction multiplier important near the disc's inner
edge. Dimensionless mass accretion rate defined as $\displaystyle \mdot =
\frac{\varkappa_T c \dot{M}}{4\pi GM}$ and midplane sonic velocity $c_{\rm s}$
(in $c$ units) may be expressed from each other as 
\begin{equation}
\mdot = 3 \alpha \tau_0 c_{\rm s}^2 r^{3/2},
\end{equation}
where $c_{\rm s}=\Omega H /c$.

Finally, the system of equations takes the form
\begin{equation}
\frac{d\tilde p_{\rm r}}{d\xi} = -\frac{9}{2}\alpha \tau_0 c_{\rm s} \frac{\varkappa}{\varkappa_{\rm T}} \tilde \rho \xi,
\end{equation}
\begin{equation}
\frac{d \tilde p}{d\xi} = - \xi \tilde \rho,
\end{equation}
\begin{equation}
\frac{d\tau}{d\xi} = -2\tau_0 \frac{\varkappa}{\varkappa_{\rm T}} \tilde \rho,
\end{equation}
where $\tilde \rho$ is found as a function of total and radiation pressure as
\begin{equation}
\tilde \rho = \frac{m c^2}{k} \left(\frac{2\sigma_{\rm SB} \varkappa_{\rm T}
  GM}{3c^5} \frac{1}{\tau_0 c_{\rm s}}\right)^{1/4} r^{3/8} c_{\rm s}^2
\frac{\tilde p - \tilde p_{\rm r}}{\tilde p_{\rm r}^{1/4}} \simeq 1.06\times 10^5 \left(\frac{M}{\Msun}\right)^{1/4} \tau_0^{-1/4} c_{\rm s}^{7/4} r^{3/8} \frac{\tilde p - \tilde p_{\rm r}}{\tilde p_{\rm r}^{1/4}}.
\end{equation}
Here, $m$ is the mean mass of a particle, set to $0.6$ of a proton mass that is a good approximation for a fully ionized solar metallicity plasma. Boundary conditions for the pressure may be set at the photosphere. For total pressure, from the hydrostatic equation, 
\begin{equation}
\displaystyle p(\tau=1) = \frac{GMH}{\varkappa R^3},
\end{equation}
that corresponds, for the normalized quantity, to 
\begin{equation}
\displaystyle \tilde{p}(\tau=1) = \frac{1}{2\tau_0} \frac{\varkappa_{\rm T}}{\varkappa}.
\end{equation}
At the same time, radiation pressure is determined by the local gas temperature, equal to effective temperature at $\tau\simeq 2/3$. As most of the disc has very large optical depth, the difference between $\tau=0$, $2/3$, and $1$ is insignificant. Therefore, we can set
\begin{equation}
p_r(\tau=1) = \frac{4\sigma_{\rm SB}}{3c} T_{\rm eff}^4,
\end{equation}
that corresponds, for the normalized quantity, to 
\begin{equation}
\tilde{p}_r(\tau=1) = 3\alpha c_{\rm s}.
\end{equation}
Thomson optical depth needs to be $\tau_0$ in the equatorial plane, that sets constrain on the only free parameter of the model, $\tau_0$. 

Once the density and pressure are known as functions of vertical coordinate, rotation profile may be expressed using equation~(\ref{E:oz}) and normalizations introduced in this Appendix
\begin{equation}\label{E:app:omega}
\displaystyle \pardir{\xi}{\ln \Omega} = \frac{H^2}{2\varpi^2} \frac{\tilde p}{\tilde \rho} \left( \Gamma_\rho \pardir{\xi}{\ln \tilde p} - \Gamma_p \pardir{\xi}{\ln \tilde \rho}\right) = \frac{1}{2} c_{\rm s}^2 \frac{\varpi c^2}{GM}\frac{\tilde p}{\tilde \rho} \left( \Gamma_\rho \pardir{\xi}{\ln \tilde p} - \Gamma_p \pardir{\xi}{\ln \tilde \rho}\right).
\end{equation}
Finally, after solving for the vertical structure, radial velocity may be
calculated directly using normalized equations~(\ref{E:wik}) and (\ref{E:angmo})
\begin{equation}\label{E:app:vr}
v_\varpi = 2\frac{\nu_c}{\varpi} \left( -\frac{3}{2}\left( \Gamma_\nu +
\Gamma_\rho + \frac{1}{2}\right)+\pardirsec{\xi}{\ln \Omega}\right) = 2\alpha
c_{\rm s}^2 \sqrt{\frac{\varpi c^2}{GM}} \left( -\frac{3}{2}\left(\Gamma_\nu +
\Gamma_\rho + \frac{1}{2}\right)+\pardirsec{\xi}{\ln \Omega}\right).
\end{equation}

\label{lastpage}

\end{document}